\documentclass[aps,preprint,preprintnumbers,amsmath,amssymb]{revtex4}
\usepackage{amsmath,mathrsfs,amsbsy,color,graphicx,bm,amsthm,amsfonts}
\usepackage{units}
\usepackage{bbm}
\usepackage{times}
\usepackage{dcolumn}
\usepackage{mathrsfs}
\usepackage{amsmath,amssymb,epsfig}
\usepackage{epstopdf}
\usepackage{hyperref}

\begin{document}

\title{Quantum properties of fermionic fields in multi-event horizon spacetime}

\author{Qianqian Liu$^{1}$, Shu-Min Wu$^{2}$, Cuihong Wen$^{1}$\footnote{Email: cuihongwen@hunnu.edu.cn}, and Jieci Wang$^{1}$\footnote{Email: jcwang@hunnu.edu.cn}}
\affiliation{$^1$  Department of Physics and Synergetic Innovation Center for Quantum Effects,\\Key Laboratory of Low-Dimensional Quantum Structures and Quantum Control of Ministry of Education, \\
 Hunan Normal University, Changsha 410081, China}
 \affiliation{$^2$  Department of Physics, Liaoning Normal University, Dalian 116029, China}

\begin{abstract}

We investigate the properties of quantum entanglement and mutual information in the multi-event horizon Schwarzschild-de Sitter (SdS) spacetime for massless Dirac fields.
We obtain the expression for the evolutions of the quantum state near the black hole event horizon (BEH) and cosmological event horizon (CEH) in the SdS spacetime.
Under the Nariai limit, the physically accessible  entanglement and mutual information are maximized, and the physically inaccessible correlations are zero.  With the increase in temperature of either horizon, the physically accessible correlations experience degradation. Notably, the initial state remains entangled and can be utilized in entanglement-based quantum information processing tasks, which differs form the scalar field case. Furthermore, the degradation of physically accessible correlations is more pronounced for small-mass black holes. In contrast, the physically inaccessible correlations separated by the CEH monotonically increase with the radiation temperature, and such correlations are not decisively influenced by the effect of particle creation at the BEH. Moreover, a similar phenomenon is observed for the inaccessible correlations separated by the BEH. This result
 differs from the single event spacetime, in which the physically inaccessible entanglement is a monotonic function of the Hawking temperature.

\end{abstract}
\vspace*{0.5cm}

\maketitle
\section{Introduction}\label{section1}

Over the past few decades, quantum entanglement has emerged as a focal point in diverse physics issues, including quantum information in many-body systems understanding of black hole formation \cite{Hawkingcreation}, and quantum nature of gravity \cite{QG1,QG2}. Since its first experimental verification \cite{experimental1,experimental2},
it has been established on a solid physical foundation and has been developed in various disciplines involving quantum teleportation \cite{teleportation1,teleportation2}, quantum metrology \cite{metrology1}, and quantum computation \cite{QT1,QT2}. In particular, quantum entanglement has been demonstrated to play a crucial role in resolving the information paradox problem of black holes  \cite{paradox1,paradox2,paradox3} and the creation of particles in the early universe and Rindler spacetime \cite{RQI1,RQI2}. Among these interesting issues, the most important process is the creation of particles near an event horizon \cite{RQI1,RQI2,RQI3,RQI4}, which may increase randomness. Consequently, the evaporation and heating of black holes destroy quantum entanglement and other quantum correlations in quantum states \cite{RQI3,RQI4,QIqianqian,W1,W2,noinertial1,BH2,BH3}. Consequently, particle creation is responsible for the degradation of physically accessible entanglement \cite{noinertial1,Dirac1} in a relativistic setting. Studies in this area are believed to provide insights into the evolution of quantum states in curved spacetime, information paradox problem of black holes, and geometry of the early universe.

Notably, all of these studies on how the creation of particles  influences the dynamics of quantum information near the event horizon were performed in asymptotically flat spacetime \cite{single1,single2}. However, observational studies suggest that the universe is expanding at an increasing rate, indicating a positive cosmological constant \cite{positive1, positive2, positive3, positive4,cosmological5}. Performing a study in spacetime with a positive cosmological constant $\Lambda$ not only provides us with a useful model for understanding the overall structure of isolated black holes in the current universe but also plays an important role in investigating the behavior of black holes formed during the early inflationary stage of the universe \cite{expansion1,Gibbons-Hawking}. In solutions of Einsteins equations where $\Lambda > 0$, an outer horizon known as the cosmological event horizon (CEH) exists  \cite{expansion1,different2,different3,different4,different5,Sds1}.

There are well-established methods for studying of quantum properties in single horizon spacetime \cite{RQI3,RQI4,QIqianqian,W1,W2,noinertial1,BH2,BH3};
however, considerably few research exist on the properties of classical and quantum correlations in multi-horizon spacetime.
Recently, Bhattacharya et al. \cite{positive5,Sds2} analyzed the behavior of quantum entanglement for the scalar field in the Schwarzschild-de Sitter (SdS) spacetime.
Herein, we analyzed the entanglement between two modes of a Dirac field described by particles near the event horizon in a multi-horizon spacetime.
Our first motivation comes from the fact that  the massless scalar field can only contain a very special type of particle, i.e., the spinless scalar bosons (Higgs,$\pi$-mesons).
For a more realistic Dirac field, a deeper understanding is lacking because of conceptual issues and greater complexity arising from the multi-component nature of the Dirac field. A comprehensive analysis of quantum information for Dirac fields allows a better understanding of quantum information and spacetime properties in the multi-event horizon spacetime. Second, the present work can provide a method to study the behavior of quantum information under the background of multi-horizon black holes. For instance, the obtained state can be used to explore the behaviors of quantum communications, quantum metrology, and the characteristics of quantum resources in a multi-event horizon spacetime.   Third, we intend to comprehend how the crucial sign change in Fermi-Dirac vs. Bose-Einstein distributions and the finite number of allowed states in fermionic systems due to the Pauli exclusion principle affect the degradation of entanglement produced by quantum thermal effects \cite{noinertial1,RQI2,Dirac1}.

We assumed that a two-mode maximum entangled state is initially shared by two experimenters, where Alice lies outside the black hole event (BEH) and Bob lies inside the CEH. Under the influence of the quantum thermal effect in the SdS spacetime, the initially prepared bipartite state transforms into a quadripartite state. As depicted in Fig. (\ref{figpenrose}), the present model involves four subsystems: the mode observed by Alice in subregion A, the mode observed by Bob in subregion B, and modes observed by two virtual observers, R and L, situated in other regions. The primary objective of our investigation is to explore the relation between quantum properties and quantum thermal effects at the BEH or CEH. We demonstrated that the quantum entanglement and mutual information are observer-dependent quantum information in the multi-event horizon SdS spacetime.

 This paper is organized as follows: Sec. II outlines the causal structure of the SdS spacetime, Sec. III analyzes the quantum entanglement and mutual information for the massless Dirac field in the SdS spacetime, and Sec. IV presents the conclusions. Throughout the paper, we use the conventions $\hbar=G=c=\kappa_{B}=1$.

\section{Quantum thermal effect in Schwarzschild-de Sitter Spacetime}\label{section2}

\begin{figure}[ht]
\centering
\includegraphics[width=0.7\textwidth]{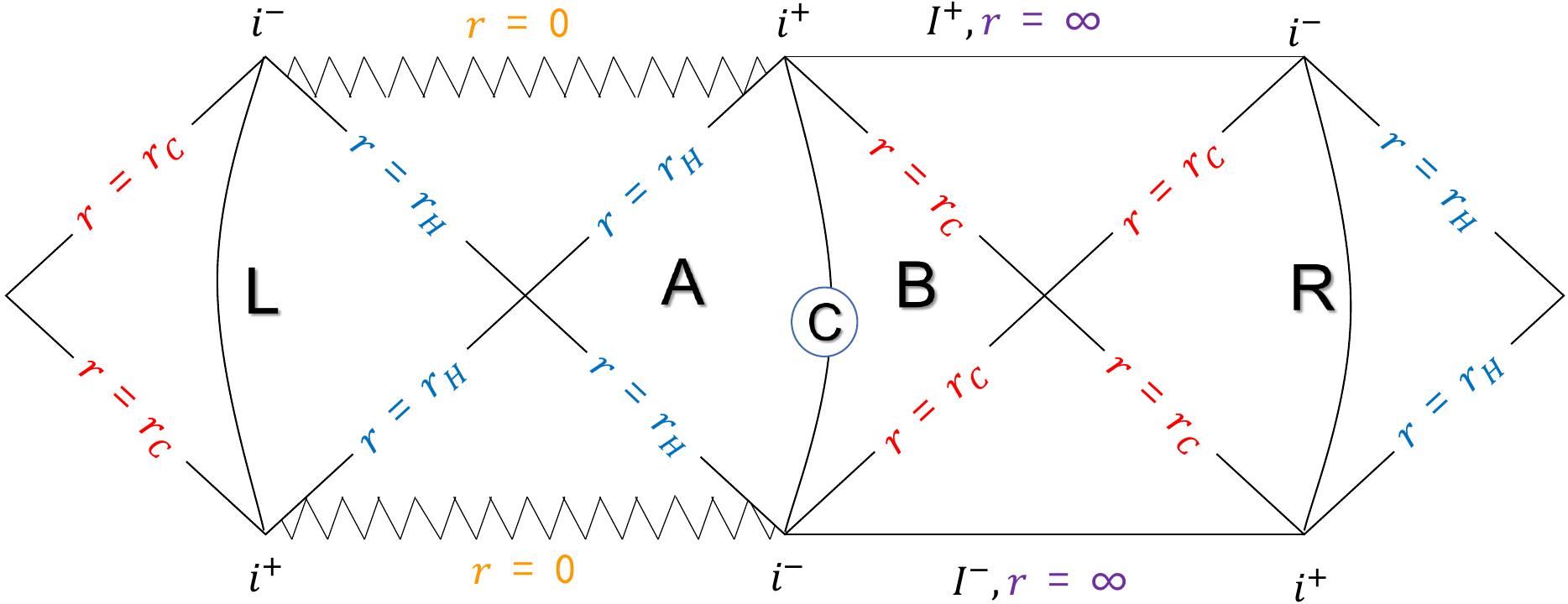}
\caption{(Color online) Penrose-Carter diagram depicts the causal structure of the extended Schwarzschild-de Sitter spacetime. The symbol $i^{\pm}$ represents the future and past time-like infinities, whereas $I^{\pm}$ represents space-like infinities. A thermally opaque membrane placed in region C divides it into subregions A and B. All the wedges in the diagram are causally disconnected.
}\label{figpenrose}
\end{figure}

Herein, the expression for the evolutions of the quantum state near the BEH and CEH in the SdS spacetime is presented. The SdS metric is given as follows \cite{Sds1,Sds2,positive5}:
\begin{equation}\label{sdsmetric}
d s^{2}=-f(r) d t^{2}+\frac{1}{f(r)} d r^{2}+r^{2}d\Omega^{2},
\end{equation}
where the metric function $f(r)$ is
\begin{equation}\label{f}
f(r)= 1-\frac{2 M}{r}-\frac{\Lambda r^{2}}{3}=\frac{\Lambda}{3r}(r-r_{H})(r_{C}-r)(r-r_{U}).
\end{equation}
In Eq. (\ref{f}), $M$ represents the black hole mass, $\Lambda>0$ represents a positive cosmological constant, $r_{C}$ and $r_{H}$ represent the locations of the cosmological and black hole horizons, respectively. The SdS metric describes a static spherically symmetric black hole in de Sitter spacetime. The Penrose figure of the SdS spacetime is depicted in Fig. (\ref{figpenrose}).

For $0<3 M \sqrt{\Lambda}<1$, the spacetime possesses BEH and CEH
	\begin{eqnarray}\label{redius}
	r_{H}&=&\frac{2}{\sqrt{\Lambda}} \cos \left[\frac{\arccos(-3 M \sqrt{\Lambda})+4\pi}{3}\right],   r_{C}=\frac{2}{\sqrt{\Lambda}} \cos \left[\frac{\arccos(-3 M \sqrt{\Lambda})}{3}\right],
	\end{eqnarray}
whereas $r_{U}=-(r_{H}+r_{C})$ is considered as an unphysical solution \cite{Sds1}.

At the Nariai limit ($3 M \sqrt{\Lambda}=1$), the two horizons coincide $r_{H}=r_{C}=1/\sqrt{\Lambda}$, corresponding to the configuration of the largest black hole and smallest de Sitter space. Unlike in single horizon spacetimes, the mass of a black hole cannot exceed a certain threshold for a given cosmological constant $\Lambda$. For $3 M \sqrt{\Lambda}>1$, the black hole horizon no longer exists, and a naked curvature singularity emerges.

The surface gravities of the BEH and CEH are given as follows:
	\begin{equation}\label{surface}
	\kappa_{H}=\frac{\Lambda\left(2 r_{H}+r_{C}\right)\left(r_{C}-r_{H}\right)}{6 r_{H}}, \quad-\kappa_{C}=\frac{\Lambda\left(2 r_{C}+r_{H}\right)\left(r_{H}-r_{C}\right)}{6 r_{C}} .
	\end{equation}
The surface gravity of the BEH is positive,
and $\kappa_{C}$ is negative because of the repulsive effects due to the cosmological constant $\Lambda>0$. In addition, the surface gravities of the two event horizons are considerably small at the Nariai limit. Therefore, this limit can be considered as the starting point for the evaporation of the black hole.

Because $r = r_{H}$, $r_{C}$ are two coordinate singularities \cite{positive5}, one can introduce the radial tortoise coordinate $r_{\star}$ represented as follows:
\begin{eqnarray}\label{tortoise2}
r_{\star}&=&\int\frac{dr}{f(r)}\\ \nonumber
&=&\frac{1}{2\kappa_H}\ln \left\vert\frac{r}{r_H}-1\right\vert -\frac{1}{2\kappa_C} \ln \left\vert1-\frac{r}{r_C}\right\vert +\frac{1}{2\kappa_U}\ln \left\vert\frac{r}{r_U}-1\right\vert.
\end{eqnarray}
 where $\kappa_U$ represents the surface gravity of the unphysical horizon $r_U$.
In contrast to single horizon spacetimes, two Kruskal coordinate patches are required to derive the nonsingular coordinate mapping of the entire SdS spacetime manifold via analytical continuation.
The SdS spacetime lacks a single Kruskal coordinate that can simultaneously remove the coordinate singularities for both horizons. Using Eq. (\ref{tortoise2}), we set two Kruskal-like coordinates to extend spacetime beyond the event horizons \cite{Sds2,positive5}
\begin{eqnarray}\label{dsh}
ds^2&=-\frac{2M}{r}\left\vert1-\frac{r}{r_C}\right\vert^{1+\frac{\kappa_H}{\kappa_C}} \left(1+\frac{r}{r_H+r_C}\right)^{1-\frac{\kappa_H}{\kappa_U}}\, d{\overline u}_H d {\overline v}_H+r^{2}d\Omega^{2},
\end{eqnarray}
\begin{eqnarray}
ds^2&=-\frac{2M}{r}\left\vert\frac{r}{r_H}-1\right\vert^{1+\frac{\kappa_C}{\kappa_H}} \left(1+\frac{r}{r_H+r_C}\right)^{1+\frac{\kappa_C}{\kappa_U}}\, d {\overline u}_C d {\overline v}_C+r^{2}d\Omega^{2},
\label{dsc}
\end{eqnarray}
 where
\begin{eqnarray}\label{newmatric}
{\overline u}_H=-\frac{1}{\kappa_H}e^{-\kappa_H u},\quad {\overline v}_H=\frac{1}{\kappa_H}e^{\kappa_H v}, \quad
{\overline u}_C=\frac{1}{\kappa_C}e^{\kappa_C u}, \textrm{and} \quad {\overline v}_C=-\frac{1}{\kappa_C}e^{-\kappa_C v}.
\end{eqnarray}
The null coordinates are defined as the usual retarded coordinate $u=t-r_{\star}$ and the advanced coordinate $v=t+r_{\star}$.
Under the limit $\Lambda\rightarrow0$ (implying $r_{C}\rightarrow\infty$, $\kappa_{C}\rightarrow0$,
$r_{H}\rightarrow2M$, and $\kappa_{H}\rightarrow1/4M$), the SdS metric has the same line element as the Schwarzschild metric.
Notably, Eqs.~(\ref{dsh}) and (\ref{dsc}) exclude the black hole and cosmological singularities, respectively. Therefore, we have a well-defined region $r_H<r<r_C$.

In general, the SdS spacetime possesses two event horizons associated with various local temperatures, rendering it difficult to achieve equilibrium. One solution to this problem is to separate the two horizons with a thermally opaque membrane \cite{positive5,Sds2,membrane1,membrane2} in the region C (C $=$ A $\cup$ B), as depicted in Fig. (\ref{figpenrose}), which prevents modes from penetrating and confines them in the respective regions. Alice, located at the black hole side, detects the Hawking radiation at temperature $\kappa_{H}/2\pi$ in subregion A, and Bob detects temperature $\kappa_{C}/2\pi$ in subregion B.

To obtain the dynamics of the Dirac field in the SdS spacetime, we considered the Dirac equation \cite{Dirac1} represented as follows:
\begin{equation}
\left[\gamma^{a} e_{a}^{\mu}\left(\partial_{\mu}+\Gamma_{\mu}\right)\right]\Psi=0,
\end{equation}
where $\gamma^{a}$ represents the Dirac matrices, the four-vectors $e_{a}^{\mu}$ represent the inverse of the tetrad $e_{\mu}^{a}$, and $\Gamma_{\mu}$ represents the spin connection coefficients.
First, we considered the side of the membrane that faces the black hole in the subregion A.
In this case, the quantization of the subregion field is similar to the black hole with a single event horizon \cite{RQI2,RQI3,RQI4}.
Specifically, one can quantize the Dirac field using the local and Kruskal modes and obtain the Bogoliubov transformations between the creation and annihilation operators in the SdS and Kruskal coordinates \cite{RQI3}. After properly normalizing the state vector, the vacuum state in the Kruskal coordinates can be expressed as follows:
\begin{equation}\label{global}
	|0\rangle_{\kappa_{H}}=\cos r\left|0_{A}, 0_{L}\right\rangle+\sin r\left|1_{A}, 1_{L}\right\rangle,
	\end{equation}
where $\tan r=e^{-\pi \omega/\kappa_{H}}$, $\left|n_{A}\right\rangle$ and $\left|n_{L}\right\rangle$ $(n\in{0,1})$ characterize the mode in region A and region L, respectively. This field quantization produces the local vacuum $|0_{A},0_{L}\rangle$ and the global vacuum $|0_{\kappa_{H}}\rangle$ in $A\cup L$ corresponds to the field quantization with the Kruskal coordinates of Eq. (\ref{dsh}). The other group is $|0_{B},0_{R}\rangle$ and the global vacuum $|0_{\kappa_{C}}\rangle$ corresponds to the field quantization of Eq. (\ref{dsc}).

In addition, we expanded the Kruskal vacuum using the internal and external modes of the CEH as follows:
\begin{equation}\label{global1}
|0\rangle_{\kappa_{C}}=\cos s\left|0_{B}, 0_{R}\right\rangle+\sin s\left|1_{B}, 1_{R}\right\rangle,
	\end{equation}
where $\tan s=e^{-\pi \omega/\kappa_{C}}$, $\left|n_{B}\right\rangle$ and $\left|n_{R}\right\rangle$ represent the mode in region B and R, respectively.
These excited states can be characterized as follows:
	\begin{equation}\label{global2}
	|1\rangle_{\kappa_{H}}=\left|1_{A}, 0_{L}\right\rangle, \quad|1\rangle_{\kappa_{C}}=\left|1_{B}, 0_{R}\right\rangle.
	\end{equation}
Due to the existence of the thermally opaque membrane \cite{positive5,Sds2,membrane1,membrane2}, the influence of one horizon can be determined by considering the other horizon as the boundary.
To achieve this, we solved the following Schr\"odinger-like equation \cite{potential,WMJ} as follows:
\begin{equation} \label{Schrodinger}
-\frac{d^2Z_\pm}{d r^{\ast 2}}+V_\pm=\omega^2 Z_\pm~,
\end{equation}
where the effective potentials $V_{\pm}$ are given by
\begin{equation}
V_\pm=W^2 \pm \frac{dW}{d r^{\ast}} = \mp K \frac{f(r) \sqrt{f(r)}}{r^2} \pm K \frac{f(r)' \sqrt{f(r)}}{2r}+ \frac{K^2 f(r)}{r^2}~,
\end{equation}
where $K$ can take positive and negative integer values
$K=\pm(l+1)$, with $l=0,1,2,...$.
The potentials $V_+$ and $V_-$ are smooth functions that vanish when approaching the $r_{H}$ and $r_{C}$ and are positive in between. Therefore, this bell-shaped potential acts as a thermally opaque membrane between the BEH and CEH.
Modes that cannot penetrate this potential will be localized near the horizons, resulting in their disconnection from each other.

\section{Quantum entanglement in Schwarzschild-de Sitter spacetime}\label{section3}
\subsection{Reduction of entanglement between initially correlated modes}
We considered a two-mode maximum entangled initial state shared by Alice and Bob with two Kruskal modes
$\kappa_{H}$ and $\kappa_{C}$ \cite{RQI2},
\begin{equation} \label{initia} |\mathrm{\psi}\rangle_{AB}=\frac{1}{\sqrt{2}}|0\rangle_{\kappa_{H}}|0\rangle_{\kappa_{C}}+\frac{1}{\sqrt{2}}|1\rangle_{\kappa_{H}}|1\rangle_{\kappa_{C }}.
	\end{equation}
The system is bipartite from the perspective of a Kruskal observer. However, as shown in Fig. (\ref{figpenrose}), two additional sets of regions, R and L, become related from the perspective of a local observer.
Using Eqs. (\ref{global}-\ref{global2}),
 the quantum state $|\mathrm{\psi}\rangle_{AB}$ evolves to
\begin{eqnarray}\label{all}
	|\mathrm{\psi}\rangle_{ALBR}&=&\frac{1}{\sqrt{2}}(\cos r\cos  s|0000\rangle+
\cos r\sin  s|0011\rangle\nonumber\\
&&+\sin r\cos  s|1100\rangle+\sin r\sin  s|1111\rangle+|1010\rangle).
	\end{eqnarray}

Because the BEH and CEH exterior regions are causally disconnected from the interior regions, only the quantum entanglement between Kruskal observers Alice and Bob is physically accessible. Considering the trace over the states belonging to the causally disconnected regions R and L, we obtained the density matrix as follows:
\begin{equation}
\hat{\rho}_{(AB)}= \frac{1}{2}\left(	\begin{array}{cccc}
\cos^{2}r \cos^{2}s &  0 &  0 & \cos r\cos s \\
		0&  \cos^{2}r\sin^{2}s  & 0 & 0 \\
		0& 0 & \sin ^{2}r\cos^{2} s  & 0 \\
\cos r\cos s&  0&  0&
 \sin^{2}r\sin^{2}s+1
	\end{array}\right).
		\end{equation}

\begin{figure}[ht]
\centering
\includegraphics[width=0.497\textwidth]{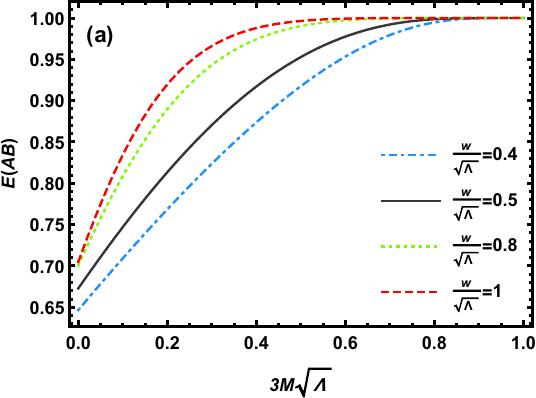}
\includegraphics[width=0.49\textwidth]{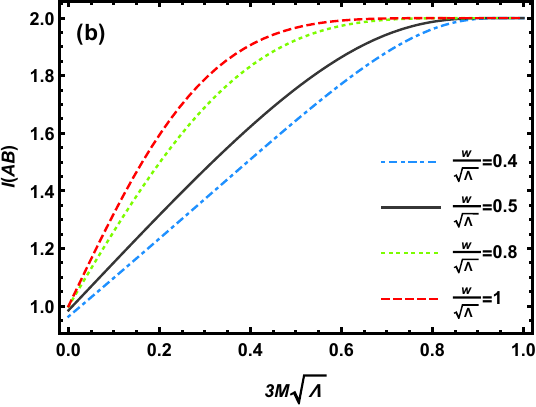}
\caption{(Color online)
 Physically accessible entanglement (a) and mutual information (b) of $\hat{\rho}_{(AB)}$ as a function of the dimensionless parameter $3M\sqrt{\Lambda}$.
}\label{fig1}
\end{figure}

In this research, we used the concurrence \cite{concurrence1,concurrence2} as an entanglement measure to investigate the quantum entanglement in different modes. For a mixed two-qubit state, the concurrence is given as follows:
\begin{eqnarray}  \label{concurrence}
C(\rho) =\max \left\{ 0,\lambda _{1}-\lambda
_{2}-\lambda _{3}-\lambda _{4}\right\}, \quad\lambda_i\ge
\lambda_{i+1}\ge 0,
\end{eqnarray}
where $\lambda_i$ represents the square roots of the eigenvalues of the
matrix $\rho\tilde{\rho}$,
$\tilde{\rho}=(\sigma_y\otimes\sigma_y)\,
\rho^{*}\,(\sigma_y\otimes\sigma_y)$ represents the ``spin-flip" matrix.
To further explore this model, we used the mutual information \cite{mutualinformation} to measure the total correlation between any two subsystems as follows:
\begin{equation}\label{mutualinformation}
\mathcal{I}_{\alpha\beta}=S\left(\rho_\alpha\right)+S\left(\rho_\beta\right)-S\left(\rho_{\alpha \beta}\right),
\end{equation}
where $S=-\operatorname{Tr}(\rho \ln \rho)$ represents the von Neuman entropy.
Furthermore, we obtained the quantum entanglement and mutual information for the state $\hat{\rho}_{(AB)}$ and plotted them in Fig. (\ref{fig1}).

Fig. (\ref{fig1}a) depicts the monotonic decrease of the physically accessible entanglement $C(\rho_{AB})$ as a function of the dimensionless parameter $3M\sqrt{\Lambda}$. At the Nariai limit, the physically accessible entanglement is found to be at its peak.  Due to the coincidence of the two horizons, the temperature of the horizons is vanishingly small at this limit. For smaller values of $\omega/\sqrt{\Lambda}$, less entanglement of state $\hat{\rho}_{(AB)}$ maintained for a given $3M\sqrt{\Lambda}$. This is because smaller values of $\omega/\sqrt{\Lambda}$ correspond to higher values of $\Lambda$ for a fixed $\omega$. Moreover, a higher value of $\Lambda$ leads to a higher rate of cosmological particle creation and a higher temperature of the CEH, eventually leading to the degradation of quantum entanglement. Furthermore, assuming a fixed cosmological constant $\Lambda$, the entanglement increases monotonically with increasing $M$ and saturates at unity in the Nariai limit. Consequently, the degradation of physically accessible entanglement is more pronounced for small-mass black holes.

This result can also be interpreted in terms of the temperature of the quantum thermal effect. The values of surface gravities of the BEH and CEH decrease with the increasing mass parameter $M$ for a fixed $\Lambda$.
If the mass parameter $M$ approaches zero, the temperature of the BEH becomes very high, leading to a high generation rate of particles at the BEH.
 Combining the above analysis, we can conclude that the physically accessible entanglement decreases with increasing temperature for both horizons.
Notably, the physically accessible entanglement in Refs. \cite{positive5,Sds2} approaches zero at limit $3M\sqrt{\Lambda}\rightarrow0$, whereas the physically accessible entanglement of the Dirac field eventually decays to a non-zero minimum at the same limit. In other words, the initial state $\hat{\rho}_{(AB)}$ remains entangled and can be utilized in entanglement-based quantum information processing tasks such as quantum teleportation and quantum metrology in the SdS spacetime.
Fig. (\ref{fig1}b) illustrates that the mutual information of the quantum state $\hat{\rho}_{(AB)}$ aligns with the qualitative conclusions drawn from the quantum entanglement analysis, implying that the particle creation at the two event horizon destroys both the initial entanglement and classic correlations.

\subsection{Generation of entanglement between initially uncorrelated modes}

The massless scalar field in Refs.\cite{positive5,Sds2} showed the entanglement between two event horizons without considering the effect of a single horizon.
On the one side, Alice cannot access the field patterns in the causally disconnected region L due to the presence of the BEH.
On the other side, Bob cannot access the field patterns in region B due to the presence of the CEH. Therefore, by tracing the states belonging to the regions B and L, we obtained the reduced density matrix separated by CEH
\begin{equation}\label{aR}
\hat{\rho}_{(AR)}= \frac{1}{2}\left(	\begin{array}{cccc}
 \cos^{2}r \cos^{2}s &  0 &  0 & 0 \\
0&  \cos^{2}r\sin^{2}s  & \sin s\cos r & 0 \\
0& \sin s\cos r & \sin^{2}r\cos^{2}s+1  & 0 \\
	0&  0&  0&\sin^{2}r\sin^{2}s
	\end{array}\right).
	 \end{equation}
We could determine the quantum entanglement and mutual information of the state $\hat{\rho}_{(AR)}$ using Eqs. (\ref{concurrence}) and (\ref{aR}).

\begin{figure}[ht]
\centering
\includegraphics[width=0.497\textwidth]{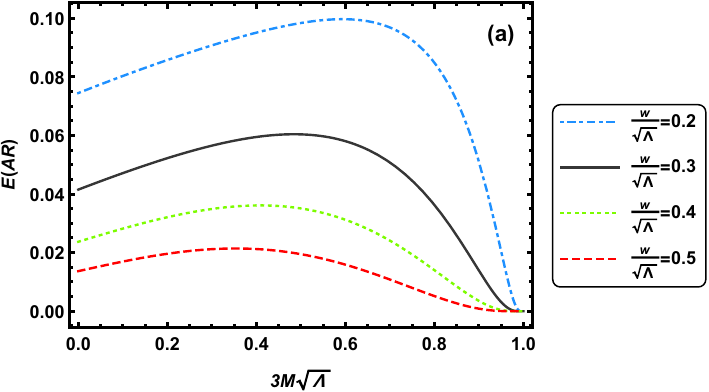}
\includegraphics[width=0.497\textwidth]{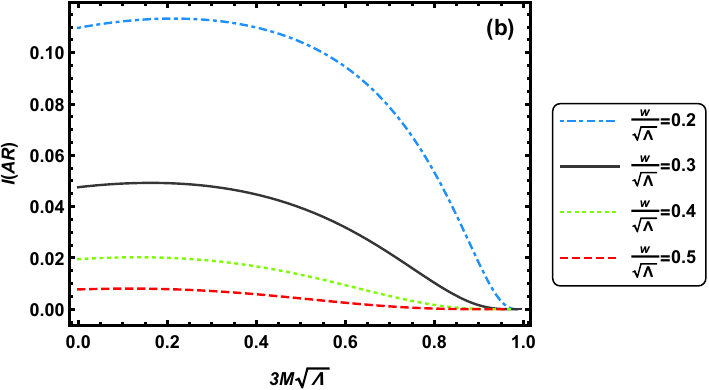}
\caption{(Color online) (a) Physically inaccessible entanglement of $\hat{\rho}_{(AR)}$ with respect to the dimensionless
 parameter $3M\sqrt{\Lambda}$.
(b) The mutual information of $\hat{\rho}_{(AR)}$ corresponding to the dimensionless parameter $3M\sqrt{\Lambda}$.
}\label{fig3}
\end{figure}

In Fig. (\ref{fig3}), we plotted the physically inaccessible entanglement $C(\rho_{AR})$ and the mutual information $I(\rho_{AR})$ as a function of the dimensionless parameter $3M\sqrt{\Lambda}$ for different values of the dimensionless parameter $\omega/\sqrt{\Lambda}$.
We found the physically inaccessible entanglement emerges at the Nariai limit, reaches a maximum value, and then decreases with decreasing $M$.
Utilizing the correspondence between the surface gravity and the Hawking temperature, it is significant to note that the mass parameter $M$ is a decreasing temperature function for both horizons, with $T_{H}$ being greater than $T_{C}$.
Then, the temperature of the BEH can either enhance or degrade the quantum entanglement between the regions A and R.
 However, for smaller values of $\omega/\sqrt{\Lambda}$, the curves exhibit higher entanglement values at a given $3M\sqrt{\Lambda}$.
The result indicates that the entanglement $C(\rho_{AR})$ is proportional to the larger cosmological constant $\Lambda$ (for a fixed $\omega$), implying that the increasing temperature and particle creation by the CEH can increase the physically inaccessible entanglement $C(\rho_{AR})$.
This finding reflects the influence of individual event horizons on quantum resources in multi-event horizon spacetimes.

Tracing over the states belonging to the regions A and R, we obtained the reduced density matrix
\begin{equation}\label{LB}
\hat{\rho}_{(LB)}= \frac{1}{2}\left(	\begin{array}{cccc}
 \cos^{2}r \cos^{2}s &  0 &  0 & 0 \\
0& \cos^{2}r\sin^{2}s+1  &
\sin r\cos s & 0 \\
0& \sin r\cos s & \sin^{2}r\cos^{2}s  & 0 \\
	0&  0&  0&\sin^{2}r\sin^{2}s
	\end{array}\right).
	 \end{equation}
Using Eqs. (\ref{concurrence}) and (\ref{LB}), we can also calculate the entanglement and mutual information for the state $\hat{\rho}_{(LB)}$.

\begin{figure}[ht]
\centering
\includegraphics[width=0.497\textwidth]{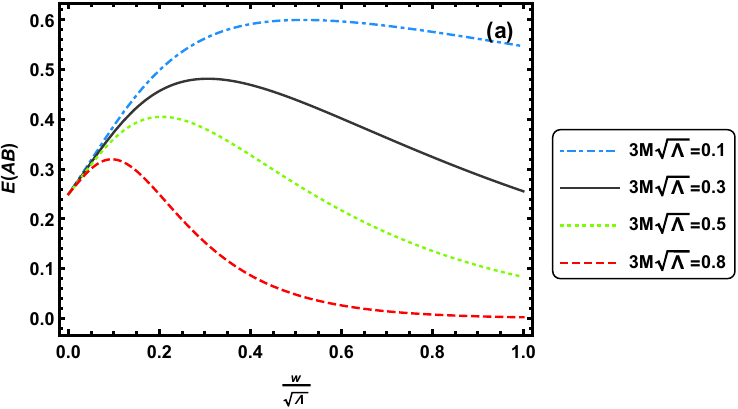}
\includegraphics[width=0.497\textwidth]{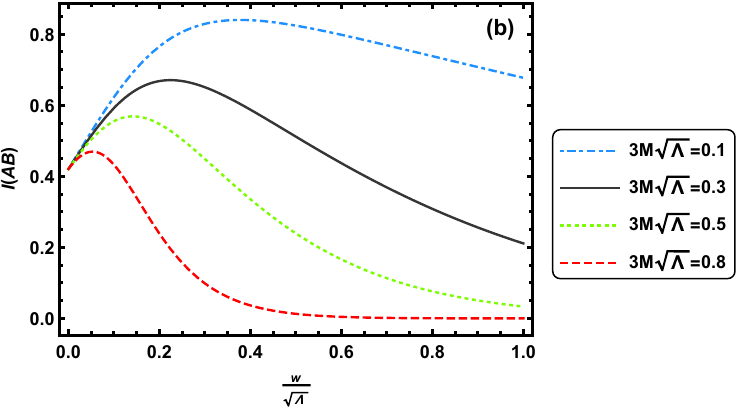}
\caption{(Color online) Physically inaccessible entanglement $C(\rho_{LB})$ (a) and mutual information $I(\rho_{LB})$ (b) with respect to the dimensionless parameter $\omega/\sqrt{\Lambda}$.
}\label{fig4}
\end{figure}

 Fig. (\ref{fig4}a) depicts the variation of the  physically inaccessible entanglement $C(\rho_{LB})$ concerning the dimensionless parameter $\omega/\sqrt{\Lambda}$ for different values of the dimensionless parameter
 $3M\sqrt{\Lambda}$.
In the previous analysis, a smaller $\omega/\sqrt{\Lambda}$ corresponded to the degradation of the physically accessible entanglement and an increase in the physically inaccessible entanglement $C(\rho_{AR})$.
However, the state $\hat{\rho}_{(LB)}$ differs from the other two quantum states in that the entanglement is no longer monotonic concerning the dimensionless parameter $\omega/\sqrt{\Lambda}$.
By fixing the cosmological constant $\Lambda$ to an effective value, the physically inaccessible entanglement $\hat{\rho}_{(LB)}$ increases with decreasing $M$ value at a given $\omega/\sqrt{\Lambda}$, corresponding to the
increasing temperature and particle creation by the BEH.
These results indicate that the physically inaccessible entanglement $C(\rho_{LB})$ increases with the temperature particle creation by the BEH, whereas the rate of cosmological particle creation has no discernible influence.
The mutual information depicted in Fig. (\ref{fig4}b) also confirms the above conclusions.

\section{Conclusions} \label{section4}

This research investigated the properties of quantum information in the multi-event horizon spacetime. The primary difference between the SdS spacetime and a single horizon spacetime is the presence of an additional CEH, leading to two distinct temperatures. We expressed evolutions of the quantum state near the BEH and CEH in the SdS spacetime. The physically accessible entanglement and mutual information are found to be the maximum, and the physically inaccessible entanglement is zero at the Nariai limit. The initial state remains entangled, and the degradation of physically accessible correlations (quantum entanglement and mutual information) is more pronounced for small-mass black holes. The physically inaccessible correlations separated by the CEH increase with the temperature particle creation by the CEH, whereas the particle creation at the BEH can either enhance or degrade them. Furthermore, we demonstrated that the physically inaccessible correlations separated by the BEH increase under the quantum thermal effect induced by the BEH, whereas the rate of cosmological particle creation has no obvious effect on the correlations. These results not only reveal the behavior of quantum information under particle creation at the BEH and CEH but also effectively reflect the influence of individual event horizons on quantum resources in multi-event horizon spacetimes.
 These results emphasize the observer-dependent nature of quantum entanglement and mutual information in the multi-event horizon spacetime.

\begin{acknowledgments}
This work is supported by the National Natural Science Foundation of China under Grant No. 12122504, No. 12374408 and No. 12205133; and the Natural Science Foundation of Hunan Province with grant No. 2023JJ30384.
\end{acknowledgments}



\end{document}